\documentclass[11pt,twoside]{article}

\usepackage{asp2006}
\usepackage{natbib,graphicx,graphics}

\begin{document}
\title{Direct imaging of extrasolar planets: overview of ground and space programs}   
\author{A. Boccaletti}   
\affil{Laboratoire d'Etude Spatiale et d'Instrumentation en Astronomie, Observatoire de Paris, F-92195 Meudon, France}    

\begin{abstract} 
With the ever-growing number of exoplanets detected, the issue of characterization is becoming more and more relevant. Direct imaging is certainly the most efficient but the most challenging tool to probe the atmosphere of exoplanets and hence in turns determine the physical properties and refine models of exoplanets. A number of instruments optimized for exoplanets imaging are now operating or planned for the short and long term both on the ground and in space. This paper reviews these instruments and their characteristics/capabilities. Conclusions are drawn on the spectral characterization point of view. 
\end{abstract}

\section{Context} 
The study of extrasolar planets has became in a decade an exciting field in modern astronomy.  
With more than 300 planets detected so far, indirect methods have been the most prolific at finding sub-stellar objects in the solar neighborhood. 
However, a few angularly resolved images of Extrasolar Giant Planets (EGPs) have been finally obtained with 8-10m class telescopes on the ground as well as in space with the HST 
\citep[for instance,][]{Chauvin05, Neuhauser05, Lafreniere08}. 
The detection rate is progressing quickly. Some discoveries were announced a week or so before the conference \citep{Kalas08, Marois08, Lagrange08}. 
These outstanding observations were made possible by the favorable configurations of these planetary systems: low star/planet mass ratio, young age (a few tens to hundreds Myr), large physical distances (typically hundreds of AU, except for $\beta$ Pic) which basically make these planets bright enough or locally above the stellar halo. A list of objects detected by direct imaging is summarized in Tab. 1.

This paper presents an overview of direct imaging projects on single apertures. As guideline for non-specialists, the discussion remains general and more details can be found in reference papers. The first part briefly reminds the problematic of high contrast imaging and the instrumental solutions that are being developed. In section \ref{sec:planned} and \ref{sec:future} the projects are separated in two categories: those that are planned and those that are proposed for the future and mostly focused on telluric planets. The intent is to clearly points out what is going to be achieved in terms of parameter space in order to better define what will be needed for the next projects. Finally, all these projects are sorted along a timeline and some conclusions are derived.

\begin{table}[!ht] 
\caption{List of planetary mass objects detected with direct imaging, as of Feb. 2009. Error bars are not indicated but are often larger than 1\,M$_J$ therefore placing some of these objects in the Brown Dwarf regime.} 
\smallskip 
\begin{center} 
{\small 
\begin{tabular}{lccl} 
\tableline 
\noalign{\smallskip} 
object 	&	estimated mass &  estimated separation & reference \\ 
		& 		[M$_J$]	 &	[AU] 				& \\
\noalign{\smallskip} 
\tableline 
\noalign{\smallskip} 
2M1207 b		& 	5 		&	46   & \citet{Chauvin05} \\ 
GQ Lup b		& 	17 		& 	100 & \citet{Neuhauser05}\\ 
AB Pic b		& 	14 		&	248 & \citet{Chauvin05b}\\ 
CHRX73 b	& 	12 		&	210 & \citet{Luhman06}\\ 
HN Peg b		& 	16 		&	795 & \citet{Luhman07} \\
DH Tau b		& 	12 		&	330 & \citet{Itoh05}\\\ 
RSX 1609 b	& 	8		&	330 & \citet{Lafreniere08} \\ 
Fomalhaut b	& 	3  		&	120 & \citet{Kalas08}\\ 
HR8799 b		& 	5 		& 	46   & \citet{Marois08}\\ 
HR8799 c		& 	12 		& 	330 & \citet{Marois08}\\ 
HR8799 d		& 	8		&  	330 & \citet{Marois08}\\ 
$\beta$ Pic b	&	8		&	8      & \citet{Lagrange08} \\
\noalign{\smallskip} 
\tableline 
\end{tabular} 
}
\end{center} 
\end{table}

\section{Problematic and solutions} 
\label{sec:problem}
The problematic is well known: planets are much fainter than parent stars and angularly close. The Sun-Jupiter example is often cited as a reference, with a contrast of about $10^9$ and a angular separation of 0.5" at 10pc. However, the realm of planets exhibits a much wider variety. The star to planet brightness ratio strongly depends on the characteristics of the system (age, temperature, physical distance, radius) and the spectral range. Actual contrasts of planets range between $10^4$ and $10^{10}$.

The huge contrast issue is emphasized by the diffraction which makes the stellar light extending all across the focal plane. As a consequence, the starlight overshines the planet light and produces a large photon noise. A system to remove or attenuate the starlight is therefore mandatory to improve the signal to noise ratio at the planet position in the field. On single aperture the solution is to use a coronagraphic system as Lyot did in the 1930's to observe the solar corona \citep{lyot39}. But, the big difference with solar observations is that stars are point like source and therefore images are dominated by diffraction. Since 1996 \citep{Gay96} alternative concepts have been studied. Large rejections have been searched for and achromaticity have been considered as a major issue. To date, many coronagraph concepts do exist. An almost exhautive list is given in \citet{guyon06} and a classification in \citet{quirrenbach05}. Also, many of them were prototyped and tested successfully at large contrast \citep[for instance]{baudoz07}. A few were also implemented on current telescopes \citep{boccaletti04}. The interesting point in this race towards high contrast is that many concepts of coronagraph provide a perfect attenuation of the starlight in some particular conditions (shape of pupil, bandwidth, ...), at least on the paper. Prototyping activities have demonstrated that the perfect starlight rejection can be approached but never reached. Nevertheless, some concepts have been elaborated to a sufficiently high level that is compatible with planet detection.  

The second issue, is related to wavefront aberrations. Even a perfect coronagraph only attenuates the coherent part of the wavefront. Aberrations are leaking through coronagraphs and produce a residual intensity in the focal plane in the form of a speckled halo. Two families of techniques are considered to tackle the speckle issue. The first one is more straightforward and is using wavefront correction and therefore necessitates a measurement of this wavefront and one or more corrective elements (usually a deformable mirror). Several testbeds are developed to address this problem of wavefront correction. Technical implementations mostly differ by the algorithm used to measure the wavefront like speckle minimization for instance \citep{borde06,giveon05}, while all testbeds consider measurements in the coronagraphic image plane and deformable mirror to apply the correction. Very high contrasts have been already demonstrated in the lab with wavefront correction \citep{trauger07}. Other designs are being studied \citep{galicher08, codona04}. 

An alternative to wavefront correction is speckle calibration, the idea being to disentangle the stellar speckles from the planet peak owing to particular characteristics. For instance, planets have spectral or polarimetric signatures not present in the starlight. On the ground, simultaneous spectral \citep{racine99} or polarimetric measurements \citep{kuhn01} are required to get rid of the atmospheric residual speckles as of static speckles. The so-called differential imaging techniques might be affected by defects inherent to the method (like chromatism for spectral differential imaging) or to the instrument (differential aberrations in the optical path). Some algorithms and optical concepts are studied to reduce the impact of these defects like spectral deconvolution \citep{sparks02}. In this respect, Integral Field Spectroscopy is considered as a promising method. 
Angular differential imaging \citep{Marois06} is another sort of speckle calibration technique. With an alt-az ground-based telescope the planet rotates with the field around the star while speckles are either static (if originating to the instrument) or rapidly evolving (if originating from the atmosphere). Aberrations that slowly evolve as the field rotates can still mimic angularly separated companions. Also, the performance is improving with the angular separation while a minimal angle can be defined (a few tenths of arcsec). This technique has recently been successful at finding 3 planets around the same star \citep{Marois08}. 
Finally, planets and stars provide incoherent wavefronts and this characteristic can also be exploited to reveal planets amongst speckles \citep{guyon04,baudoz06, codona04}.  

 The bottom line, is that coronagraph as speckle rejection concepts have been demonstrated. The ability to make these systems efficient enough is now an engineering issue. 

\section{Planned projects} 
\label{sec:planned} 
\subsection{"Planet finders" on 8-m class telescopes (2011)} 
Ground based 8-m class telescopes are now equipped with Adaptive Optics (AO) systems and some are already including high contrast imaging facilities like coronagraphy \citep{boccaletti04} or differential imaging \citep{lenzen04}. A few discoveries of giant planets were already possible as mentioned  above in some favorable conditions. To routinely achieve higher contrasts, "planet finder" instruments were planned since 2001-2002 at VLT, Gemini and Subaru. SPHERE \citep[Spectro Polarimetric High contrast Exoplanet  REsearch,][]{beuzit06}, GPI \citep[Gemini Planet Imager,][]{macintosh06} and HiCiAO \citep[High-contrast Coronagraphic Imager with Adaptive Optics,][]{tamura06} are sharing the same conceptual design which is to combine extreme AO, broad band coronagraphy and spectral differential imaging. These instruments slightly differ on the choice of the coronagraphic or differential systems but will certainly deliver very similar performance. SPHERE also extends to the visible range with a differential polarimetric imager to take advantage of the increase of the reflected light for very close planets \citep{schmid06}. 

Typical targets by order of importance are :
\begin{itemize}
\item[-] { Young and very young stars: planets in young systems are warm and hence self luminous.  Evolutionnary models \citep{burrows97, chabrier00} are predicting higher luminosity than for mature planets by several orders of magnitude depending on age and mass. Adequate targets are a few tens of Myr old and within 100\,pc. Spectral differential imaging will be efficient for planets with  $300<T (K)<1300$.}
\item[-] { Stars with known planets: the improvement of temporal coverage makes possible the detection of long period giants with radial velocity (about 100 planets with P$>$5 years according to http://exoplanet.eu). Many stars also exhibits long-term residuals indicating massive planets at large separations appropriate for direct imaging.}
\item[-] { Nearby stars ($<$5\,pc): the spectrum of irradiated giant planets at $<$1\,AU has a significant reflected light component that can be detected in the visible owing to the gain in angular resolution. }
\end{itemize}

Large surveys will be necessary with such instruments to discover new planets. Near IR spectra of these planets at low or medium resolution ($50<R<800$) will be feasible for the first time. Further details on performance estimate can be found in \citet{Marois08b,Vigan08,Thalman08,Boccaletti08}.

\subsection{JWST (2014)} 
The James Webb Space Telescope is an observatory that allows diffraction limited imaging at wavelengths longer than 2$\mu$m. It comes with a suite of 4 instruments out of 3 being equipped with coronagraphs or high contrast facilities designed for the characterization of extrasolar giant planets. 

\begin{itemize}
\item[-] {NIRCAM, the near IR camera (0.6-5$\mu$m) has 5 coronagraphs \citep{krist07} based on the band-limited concept \citep{kuchner02} and combined with 5-10\% bandwidth filters (between 2 and 4.8$\mu$m). The best performance are expected at 4.6$\mu$m owing to a peak in the thermal emission of EGPs. At this band the coronagraph is able to look at separation larger than 0.6" and will achieve a contrast level of about $10^{5}-10^{6}$ (at about 1-2"). This limit of detection \citep{krist07} corresponds to a mass of 2\,M$_J$ for 1\,Gyr and less than 1\,M$_J$ for young objects ($<$300\,Myr).}
\item[-] {MIRI, the mid IR camera (5-28$\mu$m) has 4 coronagraphs \citep{boccaletti05} based on the Four Quadrant Phase Mask concept \citep{rouan00} and combined with 5\% bandwidth filters (at 10.65, 11.40 and 15.50$\mu$m). Contrasts of  $10^{4}-10^{5}$ are achievable between 0.5 and 1". This should allow detection of 5\,M$_J$ mature planets.}
\item[-] {TFI is a tunable filter imager (1.6-4.9$\mu$m) implemented in the fine guidance sensor capable of coronagraphic imaging with a resolution of R=100 \citep{doyon08}. Taking advantage of the spectral deconvolution technique it will be able to achieve a contrast of 10$^5$ at 1" with a 0.6" lyot mask. Therefore, TFI will have similar performance has NIRCAM but with a higher spectral resolution appropriate for a finer characterization of brightest objects. }
\end{itemize}

\subsection{Extremely Large Telescopes ($>$2017)} 
The ELTs instrumentation for extrasolar planet direct imaging is more prospective, but still it is considered as one of the first priority. Two spectral regimes are considered : 
\begin{itemize}
\item[-] {EPICS \citep[Exoplanet Imaging Camera and Spectrograph for the European ELT,][]{kasper08} and PFI \citep[Planet Finder Imager of the Thirty Meter Telescope,][]{vasisht06} will provide near IR high contrast capability with unprecedented angular resolution (about 10 mas). These two instruments aim at contrasts of $10^{8}-10^{9}$ as close as 30mas to characterize mature EGPs and possibly large telluric ones. }
\item[-] {METIS \citep[Mid-infrared ELT Imager and Spectrograph,][]{brandl08} has an interesting opportunity to image irradiated planets in the mid IR (3-20$\mu$m) down to a few masses of Jupiter (again thanks to the angular resolution gain).}
\end{itemize}

\begin{figure}[h]
\centerline{\includegraphics[width=13.4cm]{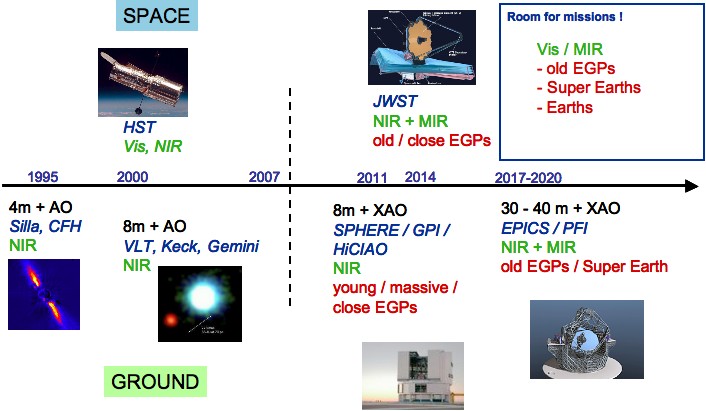}}
\caption{Timeline of direct imaging programs with a separation between ground (bottom) and space (top). Missions names, spectral range and typical detectable planets are indicated in blue, green, and red. As a result, the effort should be directed towards mature giant and telluric planets both in the visible and mid IR in order to complement  ELTs. } 
\end{figure}

\section{Future projects} 
\label{sec:future}
In contrary to section \ref{sec:planned} the following projects are not yet approved neither planned but many conceptual and technical activities are being pursued. With respect to previous projects, the objectives are focused towards lower masses (telluric planets), older  systems ($>$100\,Myr ), more distant stars ($>$100\,pc), closer planets ($<$5\,AU)  and also shorter wavelengths ($<$1$\mu$m). It is not the goal of this paper to describe all concepts. However, they can be categorized in two main families. 

The former Terrestrial Planet Finder Coronagraph was classified as a flag-ship mission. This concept has been re-baselined as a probe-class mission after 2006 with less ambitious objectives. A series of projects like PECO \citep{guyon08}, ACCESS  \citep{trauger08}, EPIC \citep{clampin07} are being studied in the US for the decadal survey. These projects assume a small (1.5-2m) telescope optimized for high-contrast imaging which necessarily means off-axis and good optical quality. Different sort of coronagraphs or wavefront correction devices are proposed to achieve $\sim10^9$ contrast. The focal instrumentation should be capable of low resolution spectroscopy (R=20-50) in the 0.4-1.0$\mu m$ range. 
In Europe, an equivalent mission, the Super Earth Explorer \citep[SEE-COAST,][]{schneider08} was submitted to Cosmic Vision in 2007 (but not selected) with polarimetric capabilities in addition to spectroscopy \citep[see][this proceeding]{baudoz09}. It was ranked as a Medium class mission according to ESA nomenclature which is equivalent to a Probe-class.
All these concepts rely for the target sample on planets detected by radial velocity surveys of which the sensitivity extends now to long periods. The prime objective is to achieve a census of giant planets on large orbits for nearby stars ($<$20-25pc) and perform spectral characterization. For the nearest stars and the closest planets accessible, large telluric planets could be observed and characterized. A high contrast imaging telescope is also suitable for the identification of circumstellar disks that are intrinsically of interest to study planetary formation and also could eventually prevent the detection of Earth-like planets with even more ambitious missions. The bottom line is to explore the diversity of planets on a pre-defined target sample.  
Also, it is recognized that the mission cost is driven by the telescope size rather than the instrument and that no major technological change would be needed for bigger telescopes. 

A second type of concept has recently emerged. The basic idea is to take advantage of Fresnel diffraction to reach higher contrast.        
One solution is to fly a large occulter in front of a telescope. A large shadow is projected at the telescope location. Performance are much less sensitive to the telescope wavefront quality but the constraints on the occulter manufacturing can be severe (of the order of 1-10mm precision). The New World Observer proposal \citep{cash08} is intended to detect telluric planets and perform characterization of narrow spectral features in the visible range ideally to pick-up biosignatures. Therefore, it requires a larger telescope than the aforesaid projects. Currently, NWO is made of a 50\,m occulter located 80000\,km in front of a 4\,m telescope. The complexity of formation fly  and deployment of large structures in space are some issues that need to be further investigated. 
An other design is proposed by \citet{Koechlin05} in a form of an Fresnel imaging lens which is supported by a spacecraft and focalizes the first diffraction order on a field telescope some kilometers away. The lens itself is an array of sub-apertures arranged in a particular form and is uses as a focuser. The image is therefore free of aberrations and could benefit to many coronagraph designs. Combined with pupil apodization, sharp PSFs can be obtained. 
The design is however chromatic and requires some corrective elements downstream. Several spectral channels would be required to achieve a large bandwidth. The NWO and Fresnel imager are more prospective concepts that need to be elaborated further in the context of space engineering. 

It is very likely that none of this concept will actually fly before 2020.

\begin{figure}[h]
\centerline{\includegraphics[width=13.4cm]{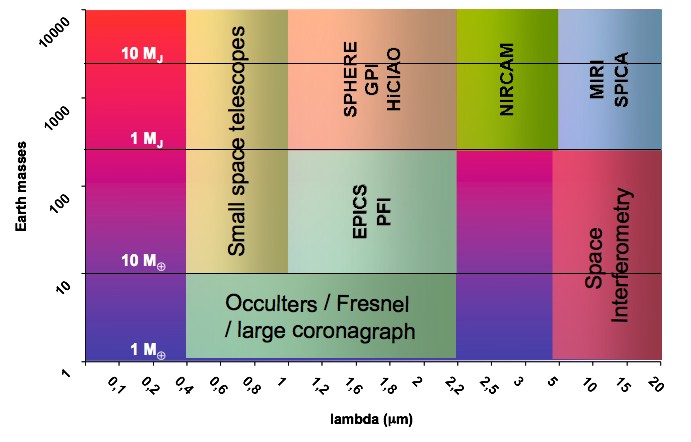}}
\caption{Diagram showing the sweetspot of each mission type as a function of wavelength.}
\end{figure}

\section{Tentative conclusions} 
From the above description, it is instructive to put on a timeline the different planned projects identifying spectral range and typical detectable planets. 
It is obviously recognized that all bandwidths are of interest for the characterization of exoplanets. For the direct imaging, this applies to visible, near IR and mid IR. What comes up from Fig. 1 is that a slot for a space mission geared toward mature giant planets either at visible or mid IR is not covered on the ground at least until 2025-2030. Another way of putting this information in a mass vs. wavelength diagram is shown in Fig. 2. If we apply a concept of minimal effort, this slot could be fill in by small telescopes optimized for a deep characterization of giant planets at visible an ideally down to  massive telluric planets that are named Super Earths. If all telluric planets share similar characteristics (atmospheric composition, geophysical attributes, biosignatures, seasonnal activities, ...), it would be worth to devote a space mission on that topic because it would be both complementary to ground-based projects and also preparatory to those that are focused to the search of life on Earth analogs \citep{cockell08}.


\bibliographystyle{pasp}
\bibliography{boccaletti_molecules}

\begin{thebibliography}{}

\bibitem[\protect\citeauthoryear{{Baudoz} et~al.}{{Baudoz}
  et~al.}{2006}]{baudoz06}
{Baudoz}, P., {Boccaletti}, A., {Baudrand}, J.,  and {Rouan}, D. 2006, in IAU
  Colloq. 200: Direct Imaging of Exoplanets: Science and Techniques, ed.
  C.~{Aime} and F.~{Vakili}, p. 553

\bibitem[\protect\citeauthoryear{{Baudoz}, {Boccaletti}, and {Rouan}}{{Baudoz}
  et~al.}{2007}]{baudoz07}
{Baudoz}, P., {Boccaletti}, A.,  and {Rouan}, D. 2007, in In the Spirit of
  Bernard Lyot: The Direct Detection of Planets and Circumstellar Disks in the
  21st Century, ed. P.~{Kalas}

\bibitem[\protect\citeauthoryear{{Baudoz} et~al.}{{Baudoz}
  et~al.}{2009}]{baudoz09}
{Baudoz}, P., {Schneider}, J., {Boccaletti}, A., {Galicher}, R., {Stam}, D.,
  and {Tinetti}, G. 2009, in this proceeding

\bibitem[\protect\citeauthoryear{{Beuzit} et~al.}{{Beuzit}
  et~al.}{2006}]{beuzit06}
{Beuzit}, J.-L. et~al. 2006, The Messenger, 125, 29

\bibitem[\protect\citeauthoryear{{Boccaletti} et~al.}{{Boccaletti}
  et~al.}{2005}]{boccaletti05}
{Boccaletti}, A., {Baudoz}, P., {Baudrand}, J., {Reess}, J.~M.,  and {Rouan},
  D. 2005, Advances in Space Research, 36, 1099

\bibitem[\protect\citeauthoryear{{Boccaletti} et~al.}{{Boccaletti}
  et~al.}{2008}]{Boccaletti08}
{Boccaletti}, A., {Carbillet}, M., {Fusco}, T., {Mouillet}, D., {Langlois}, M.,
  {Moutou}, C.,  and {Dohlen}, K. 2008, in Society of Photo-Optical
  Instrumentation Engineers (SPIE) Conference Series, Vol. 7015, Society of
  Photo-Optical Instrumentation Engineers (SPIE) Conference Series

\bibitem[\protect\citeauthoryear{{Boccaletti} et~al.}{{Boccaletti}
  et~al.}{2004}]{boccaletti04}
{Boccaletti}, A., {Riaud}, P., {Baudoz}, P., {Baudrand}, J., {Rouan}, D.,
  {Gratadour}, D., {Lacombe}, F.,  and {Lagrange}, A.-M. 2004, \pasp, 116, 1061

\bibitem[\protect\citeauthoryear{{Bord{\'e}} and {Traub}}{{Bord{\'e}} and
  {Traub}}{2006}]{borde06}
{Bord{\'e}}, P.~J. and {Traub}, W.~A. 2006, \apj, 638, 488

\bibitem[\protect\citeauthoryear{{Brandl} et~al.}{{Brandl}
  et~al.}{2008}]{brandl08}
{Brandl}, B.~R. et~al. 2008, in Society of Photo-Optical Instrumentation
  Engineers (SPIE) Conference Series, Vol. 7014, Society of Photo-Optical
  Instrumentation Engineers (SPIE) Conference Series

\bibitem[\protect\citeauthoryear{{Burrows} et~al.}{{Burrows}
  et~al.}{1997}]{burrows97}
{Burrows}, A. et~al. 1997, \apj, 491, 856

\bibitem[\protect\citeauthoryear{{Cash} et~al.}{{Cash} et~al.}{2008}]{cash08}
{Cash}, W., {Oakley}, P., {Turnbull}, M., {Glassman}, T., {Lo}, A., {Polidan},
  R., {Kilston}, S.,  and {Noecker}, C. 2008, in Society of Photo-Optical
  Instrumentation Engineers (SPIE) Conference Series, Vol. 7010, Society of
  Photo-Optical Instrumentation Engineers (SPIE) Conference Series

\bibitem[\protect\citeauthoryear{{Chabrier} et~al.}{{Chabrier}
  et~al.}{2000}]{chabrier00}
{Chabrier}, G., {Baraffe}, I., {Allard}, F.,  and {Hauschildt}, P. 2000, \apj,
  542, 464

\bibitem[\protect\citeauthoryear{{Chauvin} et~al.}{{Chauvin}
  et~al.}{2005a}]{Chauvin05}
{Chauvin}, G., {Lagrange}, A.-M., {Dumas}, C., {Zuckerman}, B., {Mouillet}, D.,
  {Song}, I., {Beuzit}, J.-L.,  and {Lowrance}, P. 2005a, \aap, 438, L25

\bibitem[\protect\citeauthoryear{{Chauvin} et~al.}{{Chauvin}
  et~al.}{2005b}]{Chauvin05b}
{Chauvin}, G. et~al. 2005b, \aap, 438, L29

\bibitem[\protect\citeauthoryear{{Clampin}}{{Clampin}}{2007}]{clampin07}
{Clampin}, M. 2007, in In the Spirit of Bernard Lyot: The Direct Detection of
  Planets and Circumstellar Disks in the 21st Century, ed. P.~{Kalas}

\bibitem[\protect\citeauthoryear{{Cockell} et~al.}{{Cockell}
  et~al.}{2008}]{cockell08}
{Cockell}, C.~S. et~al. 2008, Experimental Astronomy, 46

\bibitem[\protect\citeauthoryear{{Codona} and {Angel}}{{Codona} and
  {Angel}}{2004}]{codona04}
{Codona}, J.~L. and {Angel}, R. 2004, \apjl, 604, L117

\bibitem[\protect\citeauthoryear{{Doyon} et~al.}{{Doyon}
  et~al.}{2008}]{doyon08}
{Doyon}, R. et~al. 2008, in Society of Photo-Optical Instrumentation Engineers
  (SPIE) Conference Series, Vol. 7010, Society of Photo-Optical Instrumentation
  Engineers (SPIE) Conference Series

\bibitem[\protect\citeauthoryear{{Galicher}, {Baudoz}, and
  {Rousset}}{{Galicher} et~al.}{2008}]{galicher08}
{Galicher}, R., {Baudoz}, P.,  and {Rousset}, G. 2008, \aap, 488, L9

\bibitem[\protect\citeauthoryear{{Gay} and {Rabbia}}{{Gay} and
  {Rabbia}}{1996}]{Gay96}
{Gay}, J. and {Rabbia}, Y. 1996, Academie des Science Paris Comptes Rendus
  Serie B Sciences Physiques, 322, 265

\bibitem[\protect\citeauthoryear{{Give'on} et~al.}{{Give'on}
  et~al.}{2005}]{giveon05}
{Give'on}, A., {Kasdin}, N.~J., {Vanderbei}, R.~J.,  and {Avitzour}, Y. 2005,
  in Society of Photo-Optical Instrumentation Engineers (SPIE) Conference
  Series, Vol. 5905, Society of Photo-Optical Instrumentation Engineers (SPIE)
  Conference Series, ed. D.~R. {Coulter}, p. 368

\bibitem[\protect\citeauthoryear{{Guyon}}{{Guyon}}{2004}]{guyon04}
{Guyon}, O. 2004, \apj, 615, 562

\bibitem[\protect\citeauthoryear{{Guyon} et~al.}{{Guyon}
  et~al.}{2008}]{guyon08}
{Guyon}, O. et~al. 2008, in Society of Photo-Optical Instrumentation Engineers
  (SPIE) Conference Series, Vol. 7010, Society of Photo-Optical Instrumentation
  Engineers (SPIE) Conference Series

\bibitem[\protect\citeauthoryear{{Guyon} et~al.}{{Guyon}
  et~al.}{2006}]{guyon06}
{Guyon}, O., {Pluzhnik}, E.~A., {Kuchner}, M.~J., {Collins}, B.,  and
  {Ridgway}, S.~T. 2006, \apjs, 167, 81

\bibitem[\protect\citeauthoryear{{Itoh} et~al.}{{Itoh} et~al.}{2005}]{Itoh05}
{Itoh}, Y. et~al. 2005, \apj, 620, 984

\bibitem[\protect\citeauthoryear{{Kalas} et~al.}{{Kalas}
  et~al.}{2008}]{Kalas08}
{Kalas}, P. et~al. 2008, ArXiv e-prints

\bibitem[\protect\citeauthoryear{{Kasper} et~al.}{{Kasper}
  et~al.}{2008}]{kasper08}
{Kasper}, M.~E. et~al. 2008, in Society of Photo-Optical Instrumentation
  Engineers (SPIE) Conference Series, Vol. 7015, Society of Photo-Optical
  Instrumentation Engineers (SPIE) Conference Series

\bibitem[\protect\citeauthoryear{{Koechlin}, {Serre}, and {Duchon}}{{Koechlin}
  et~al.}{2005}]{Koechlin05}
{Koechlin}, L., {Serre}, D.,  and {Duchon}, P. 2005, \aap, 443, 709

\bibitem[\protect\citeauthoryear{{Krist}}{{Krist}}{2007}]{krist07}
{Krist}, J. 2007, in In the Spirit of Bernard Lyot: The Direct Detection of
  Planets and Circumstellar Disks in the 21st Century, ed. P.~{Kalas}

\bibitem[\protect\citeauthoryear{{Kuchner} and {Traub}}{{Kuchner} and
  {Traub}}{2002}]{kuchner02}
{Kuchner}, M.~J. and {Traub}, W.~A. 2002, \apj, 570, 900

\bibitem[\protect\citeauthoryear{{Kuhn}, {Potter}, and {Parise}}{{Kuhn}
  et~al.}{2001}]{kuhn01}
{Kuhn}, J.~R., {Potter}, D.,  and {Parise}, B. 2001, \apjl, 553, L189

\bibitem[\protect\citeauthoryear{{Lafreni{\`e}re}, {Jayawardhana}, and {van
  Kerkwijk}}{{Lafreni{\`e}re} et~al.}{2008}]{Lafreniere08}
{Lafreni{\`e}re}, D., {Jayawardhana}, R.,  and {van Kerkwijk}, M.~H. 2008,
  \apjl, 689, L153

\bibitem[\protect\citeauthoryear{{Lagrange} et~al.}{{Lagrange}
  et~al.}{2008}]{Lagrange08}
{Lagrange}, A.~. et~al. 2008, ArXiv e-prints

\bibitem[\protect\citeauthoryear{{Lenzen} et~al.}{{Lenzen}
  et~al.}{2004}]{lenzen04}
{Lenzen}, R., {Close}, L., {Brandner}, W., {Biller}, B.,  and {Hartung}, M.
  2004, in Society of Photo-Optical Instrumentation Engineers (SPIE) Conference
  Series, Vol. 5492, Society of Photo-Optical Instrumentation Engineers (SPIE)
  Conference Series, ed. A.~F.~M. {Moorwood} and M.~{Iye}, p. 970

\bibitem[\protect\citeauthoryear{{Luhman} et~al.}{{Luhman}
  et~al.}{2007}]{Luhman07}
{Luhman}, K.~L. et~al. 2007, \apj, 654, 570

\bibitem[\protect\citeauthoryear{{Luhman} et~al.}{{Luhman}
  et~al.}{2006}]{Luhman06}
{Luhman}, K.~L. et~al. 2006, \apj, 649, 894

\bibitem[\protect\citeauthoryear{{Lyot}}{{Lyot}}{1939}]{lyot39}
{Lyot}, B. 1939, \mnras, 99, 538

\bibitem[\protect\citeauthoryear{{Macintosh} et~al.}{{Macintosh}
  et~al.}{2006}]{macintosh06}
{Macintosh}, B. et~al. 2006, in SPIE Conference, Vol. 6272, Advances in
  Adaptive Optics II. Edited by Ellerbroek, Brent L.; Bonaccini Calia,
  Domenico. Proceedings of the SPIE, Volume 6272, pp. 62720L (2006).

\bibitem[\protect\citeauthoryear{{Marois} et~al.}{{Marois}
  et~al.}{2006}]{Marois06}
{Marois}, C., {Lafreni{\`e}re}, D., {Doyon}, R., {Macintosh}, B.,  and
  {Nadeau}, D. 2006, \apj, 641, 556

\bibitem[\protect\citeauthoryear{{Marois} et~al.}{{Marois}
  et~al.}{2008a}]{Marois08}
{Marois}, C., {Macintosh}, B., {Barman}, T., {Zuckerman}, B., {Song}, I.,
  {Patience}, J., {Lafreni{\`e}re}, D.,  and {Doyon}, R. 2008a, Science, 322,
  1348

\bibitem[\protect\citeauthoryear{{Marois} et~al.}{{Marois}
  et~al.}{2008b}]{Marois08b}
{Marois}, C., {Macintosh}, B., {Soummer}, R., {Poyneer}, L.,  and {Bauman}, B.
  2008b, in Society of Photo-Optical Instrumentation Engineers (SPIE)
  Conference Series, Vol. 7015, Society of Photo-Optical Instrumentation
  Engineers (SPIE) Conference Series

\bibitem[\protect\citeauthoryear{{Neuh{\"a}user} et~al.}{{Neuh{\"a}user}
  et~al.}{2005}]{Neuhauser05}
{Neuh{\"a}user}, R., {Guenther}, E.~W., {Wuchterl}, G., {Mugrauer}, M.,
  {Bedalov}, A.,  and {Hauschildt}, P.~H. 2005, \aap, 435, L13

\bibitem[\protect\citeauthoryear{{Quirrenbach}}{{Quirrenbach}}{2005}]{quirrenb%
ach05}
{Quirrenbach}, A. 2005, ArXiv Astrophysics e-prints

\bibitem[\protect\citeauthoryear{{Racine} et~al.}{{Racine}
  et~al.}{1999}]{racine99}
{Racine}, R., {Walker}, G.~A.~H., {Nadeau}, D., {Doyon}, R.,  and {Marois}, C.
  1999, \pasp, 111, 587

\bibitem[\protect\citeauthoryear{{Rouan} et~al.}{{Rouan}
  et~al.}{2000}]{rouan00}
{Rouan}, D., {Riaud}, P., {Boccaletti}, A., {Cl{\'e}net}, Y.,  and {Labeyrie},
  A. 2000, \pasp, 112, 1479

\bibitem[\protect\citeauthoryear{{Schmid} et~al.}{{Schmid}
  et~al.}{2006}]{schmid06}
{Schmid}, H.~M. et~al. 2006, in IAU Colloq. 200: Direct Imaging of Exoplanets:
  Science and Techniques, ed. C.~{Aime} and F.~{Vakili}, p. 165

\bibitem[\protect\citeauthoryear{{Schneider} et~al.}{{Schneider}
  et~al.}{2008}]{schneider08}
{Schneider}, J. et~al. 2008, ArXiv e-prints

\bibitem[\protect\citeauthoryear{{Sparks} and {Ford}}{{Sparks} and
  {Ford}}{2002}]{sparks02}
{Sparks}, W.~B. and {Ford}, H.~C. 2002, \apj, 578, 543

\bibitem[\protect\citeauthoryear{{Tamura} et~al.}{{Tamura}
  et~al.}{2006}]{tamura06}
{Tamura}, M. et~al. 2006, in Society of Photo-Optical Instrumentation Engineers
  (SPIE) Conference Series, Vol. 6269, Society of Photo-Optical Instrumentation
  Engineers (SPIE) Conference Series

\bibitem[\protect\citeauthoryear{{Thalmann} et~al.}{{Thalmann}
  et~al.}{2008}]{Thalman08}
{Thalmann}, C. et~al. 2008, in Society of Photo-Optical Instrumentation
  Engineers (SPIE) Conference Series, Vol. 7014, Society of Photo-Optical
  Instrumentation Engineers (SPIE) Conference Series

\bibitem[\protect\citeauthoryear{{Trauger} et~al.}{{Trauger}
  et~al.}{2008}]{trauger08}
{Trauger}, J. et~al. 2008, in Society of Photo-Optical Instrumentation
  Engineers (SPIE) Conference Series, Vol. 7010, Society of Photo-Optical
  Instrumentation Engineers (SPIE) Conference Series

\bibitem[\protect\citeauthoryear{{Trauger} and {Traub}}{{Trauger} and
  {Traub}}{2007}]{trauger07}
{Trauger}, J.~T. and {Traub}, W.~A. 2007, \nat, 446, 771

\bibitem[\protect\citeauthoryear{{Vasisht} et~al.}{{Vasisht}
  et~al.}{2006}]{vasisht06}
{Vasisht}, G., {Crossfield}, I.~J., {Dumont}, P.~J., {Levine}, B.~M., {Troy},
  M., {Shao}, M., {Shelton}, J.~C.,  and {Wallace}, J.~K. 2006, in Society of
  Photo-Optical Instrumentation Engineers (SPIE) Conference Series, Vol. 6272,
  Society of Photo-Optical Instrumentation Engineers (SPIE) Conference Series

\bibitem[\protect\citeauthoryear{{Vigan} et~al.}{{Vigan}
  et~al.}{2008}]{Vigan08}
{Vigan}, A., {Langlois}, M., {Moutou}, C.,  and {Dohlen}, K. 2008, \aap, 489,
  1345

\end{thebibliography}
\end{document}